\documentstyle[twocolumn,aps,prl,epsf]{revtex}

\begin{document}
\draft

 \newcommand{\dtens}{
   \makebox[1.3\width][r]{  \makebox[1.0em][l]{}%
 \makebox[-1.9em]{%
 \raisebox{1.7ex}{$\scriptscriptstyle\leftrightarrow$}}\deut%
   }
 }

\newcommand{\newsym}[1]{\ifmmode{#1}\else{$#1$}\fi}
\newcommand{\thetae}{\newsym{\theta_e}} 
\newcommand{\pid}{\newsym{P_{id}}} 
\newcommand{\taup}{\newsym{\tau_p}} 
\newcommand{\ttz}{\newsym{T_{20}}} 
\newcommand{\ttzst}{\newsym{T_{20}(70^o)}} 
\newcommand{\tto}{\newsym{T_{21}}} 
\newcommand{\ttt}{\newsym{T_{22}}} 
\newcommand{\tti}{\newsym{T_{2i}}} 
\newcommand{\tij}{\newsym{T_{ij}}} 
\newcommand{\dtz}{\newsym{d_{20}}} 
\newcommand{\dto}{\newsym{d_{21}}} 
\newcommand{\dtt}{\newsym{d_{22}}} 
\newcommand{\dti}{\newsym{d_{2i}}} 
\newcommand{\atd}{\newsym{A_d^T}} 
\newcommand{\chis}{\newsym{\chi^2}} 
\newcommand{\chisred}{\newsym{\chi^2/N}} 
\newcommand{\mevc}{MeV/$c$}
\newcommand{\fermi}{fm$^{-1}$}
\newcommand{\deut}{\mbox{\newsym{^{2}\text{H}}}}
\newcommand{\trit}{\textit{\newsym{^{3}\text{H}}}}
\newcommand{\deed}{\mbox{\deut(e,e$^\prime$d)}}
\newcommand{\dtenseed}{\textit{\dtens\hspace{-0.3em}(e,e$^\prime$d)}}
\newcommand{\tdna}{\textit{\trit(\hspace{0.2em}\dtens,\it{n})$\,\alpha$}}
\newcommand{\sfa}{\newsym{A}}
\newcommand{\sfb}{\newsym{B}}
\newcommand{\fc}{\newsym{G_C}}
\newcommand{\fq}{\newsym{G_Q}}
\newcommand{\fm}{\newsym{G_M}}
\newcommand{\Q}{\newsym{Q}}
\newcommand{\qthree}{\newsym{\vec{q}}}
\newcommand{\pzz}{\newsym{P_{zz}}}
\newcommand{\dpzz}{\newsym{\Delta P_{zz}}}
\newcommand{\limeas}{\newsym{{\mathcal{L}}_{meas}}}
\newcommand{\licalc}{\newsym{{\mathcal{L}}_{calc}}}
\newcommand{\moller}{M\o ller}
\newcommand{\updeg}{\newsym{^{o}}}
\newcommand{\etal}{\emph{et~al.}}

\title{ Measurement of  \ttz\ in 
Elastic Electron-Deuteron Scattering }

\author{
M.\ Bouwhuis,$^1$\
R.\ Alarcon,$^2$\
T.\ Botto,$^1$\
J.\ F.\ J.\ van\ den\ Brand,$^{1,3}$\
H.\ J.\ Bulten,$^3$\ 
S.\ Dolfini,$^2$\ 
R.\ Ent,$^{5,6}$\\
M.\ Ferro-Luzzi,$^{1}$\
D.\ W.\ Higinbotham,$^8$\
C.\ W.\ de\ Jager,$^{1,5,8}$\ 
J.\ Lang,$^7$\ 
D.\ J.\ J.\ de\ Lange,$^1$\\
N.\ Papadakis,$^1$\ 
I.\ Passchier,$^1$\ 
H.\ R.\ Poolman,$^1$\ 
E.\ Six,$^2$\
J.\ J.\ M.\ Steijger,$^1$\ 
N.\ Vodinas,$^1$\ 
H.\ de\ Vries,$^1$\
Z.-L.\ Zhou$^4$\ \\
}

\address{\hss\\
$^1$\ Nationaal Instituut voor Kernfysica en Hoge-Energie Fysica,
P.O. Box 41882,  1009 DB Amsterdam, The Netherlands\\
$^2$\ Department of Physics, Arizona State University, Tempe, AZ 85287, USA\\
$^3$\ Department of Physics and Astronomy, Vrije Universiteit,
1081 HV Amsterdam, The Netherlands\\
$^4$\ Department of Physics, University of Wisconsin, Madison,
WI 53706, USA\\
$^5$\ Thomas Jefferson National Accelerator Facility, Newport News, VA 23606, USA\\
$^6$\ Department of Physics, Hampton University, Hampton, VA 23668, USA\\
$^7$\ Institut f\"ur Teilchenphysik, Eidgen\"ossische Technische Hochschule, CH-8093
Z\"urich, Switzerland\\
$^8$\ Department of Physics, University of Virginia,
Charlottesville, VA 22901, USA\\
\hss\\
}

\date{\today}

\twocolumn

\maketitle

\begin{abstract}
We report on  a measurement of the tensor analyzing power 
\ttz\ in elastic electron-deuteron scattering in the range of
four-momentum transfer from 1.8 to 3.2 \fermi. 
Electrons of 704 MeV were scattered from a polarized deuterium internal
target. 
The tensor polarization of the deuterium nuclei was 
determined with an ion-extraction system, allowing an
absolute measurement of \ttz.
The data are described well by a non-relativistic calculation that includes
the effects of meson-exchange currents.
\end{abstract}


The deuteron, as the simplest nucleus, serves as a sensitive testing ground
for a variety of nuclear models (non-relativistic \cite{wiringa,mosconi},
fully covariant \cite{hummel,vanorden}).  
The charge and current distributions inside the nucleus can be probed
with elastic electron scattering at intermediate energies.
Elastic electron scattering off the  spin-1 deuteron is completely
described in terms of three electro-magnetic form factors:
the charge monopole \fc, the magnetic dipole \fm\ and the 
charge quadrupole \fq. 
Measurement of the unpolarized cross section yields
the structure functions \sfa(\fc,\fm,\fq) and \sfb(\fm). 
When the tensor analyzing power \ttz\ is also determined,    
all three  form factors can be separated \cite{donnelly}.
A large body of data is available for \sfa\ and \sfb\ for values of the
four-momentum transfer \Q\ of up to
12 \fermi, while \ttz\ has been measured up to 4
\fermi, albeit with limited accuracy. 
The observable \ttz\ contains an interference between \fc\ and \fq\ and is
thus sensitive to the effects of short-range and tensor correlations
in the ground-state wave function of the deuteron.
In this paper absolute measurements are presented on the analyzing powers
in the \dtenseed -reaction for \Q-values
between 1.8 and 3.2 \fermi\ with a high accuracy.

The cross section for elastic electron-deuteron scattering 
with unpolarized electrons and tensor-polarized deuterium nuclei
can be expressed as  \cite{donnelly}
\begin{eqnarray}\label{eq:xsection}
\sigma  &=&  \sigma_0\Big[~1 +\frac{\atd P_{zz}}{\sqrt{2}} \Big] ,
\quad {\rm with} \\
\quad \atd  &=&  \sum_{i=0}^2 d_{2i} T_{2i}  ~\quad~ 
   {\rm and} ~\quad~  d_{20}  =  \frac{3\cos ^2\theta^*-1}{2} , \nonumber \\
d_{21}  &=&  -\sqrt{\frac{3}{2}}\sin 2\theta^* \cos\phi^* , ~\quad~
   d_{22}  =  \sqrt{\frac{3}{2}} \sin ^2\theta^* \cos 2\phi^* , \nonumber
\end{eqnarray}
with $\sigma_0$ the unpolarized cross section, $T_{2i}$ the tensor analyzing
powers and $P_{zz}$ the degree of
tensor polarization. The polarization axis of the deuteron is
defined by the angles $\theta^*$ and $\phi^*$  in the frame where the
z-axis is along the direction of the three-momentum transfer \qthree\ and
the x-axis is perpendicular to z in the scattering plane.

The experiment was performed using a 704 MeV electron beam in the AmPS
storage ring \cite{luijckx} and a tensor-polarized deuterium internal
target \cite{zhou} at NIKHEF.
By stacking several pulses of electrons,
produced by the medium-energy accelerator, circulating currents of up to
150~mA were stored in the ring.  A beam life time in excess of 2000~s was
obtained by compensating synchrotron radiation losses with a 476~MHz
cavity. 

Nuclear-polarized deuterium gas was provided by an atomic beam source. 
Deuterium atoms are produced
by means of an RF dissociator. Atoms with their electron spin up are focused
into the  target-cell feed tube by two sextupole magnets, whereas those
with spin down are defocused. A medium- and a strong-field
RF-unit induce transitions between the hyperfine states, resulting in a
tensor polarization $P_{zz}^-$ ($P_{zz}^+$) of ideally -2 (+1) with zero vector 
polarization. The tensor polarization was flipped every 20 s between
$P_{zz}^-$ and $P_{zz}^+$. The atomic beam is fed into an open-ended T-shaped
dwell cell with a diameter of 15~mm and a length of 400~mm. The cell was
cooled to approximately 150 K.  With a flux of $1.3~\times~10^{16}$ atoms/s in two
hyperfine states into the cell an integrated
target density was obtained of $2~\times~10^{13}$ atoms/cm$^2$. The
direction  of the deuteron polarization axis was defined by a magnetic
holding field (B=23~mT) and chosen to be parallel to the
average three-momentum transfer. 

Two polarimeters were available to study the polarization in the dwell
cell. A small sample (10~\%) of the atomic beam was continuously analyzed
by a Breit-Rabi polarimeter. The nuclear  polarization of the atoms
and the composition of the gas in the dwell cell was
measured with an ion-extraction system \cite{zhou1}. 
Ions, produced by the circulating electrons, were extracted from the beam line
and transported through a Wien filter (an $E \times B$ velocity selector).
Since molecular and atomic deuterium ions have different velocities,
measuring the ion current as a function of the Wien filter $B$ field allows
determination of the  atomic fraction averaged over the target cell. 
The nuclear polarization 
can be determined by accelerating the ions onto a tritium target and using
the  well-known analyzing power \cite{ohl71} of the low-energy reaction \tdna. 
We measured the polarization of molecules, originating from recombination in
the cell, in a dedicated experiment \cite{brand}. Combining these
measurements the effective target polarization was determined  to be
$\dpzz = P_{zz}^+ - P_{zz}^- =  1.175 \pm 0.057$.

The scattered electrons were detected in an electromagnetic calorimeter
\cite{passchier} consisting of 6 layers of CsI(Tl) crystals with a total
depth of 19 radiation lengths. The first layer of CsI(Tl) was sandwiched
between two plastic scintillators. The second of these, shielded from low-energy 
\moller\ electrons, provided  the trigger. A pair of wire chambers provides
tracking information of the detected electrons. The calorimeter, with
an acceptance of approximately 150 msr,  was
positioned at a central angle $\theta_e$ of 45\updeg. 

The ejected or recoiling hadrons were detected in coincidence in a so-called range
telescope (RT) \cite{brink}, consisting of 16 layers of plastic scintillator. 
The first layer had a thickness of 2~mm, all following layers were 10~mm
thick. This detector was also preceded by two wire chambers, and was positioned
at a central angle of 62.3\updeg. 
The kinetic energy of the recoiling deuterons  was kinematically
limited to 120~MeV. 

Event selection was based on  coincidence timing between the two
arms, the response of the RT scintillators and on tracking information.  
The coincidence time was corrected for effects from walk, time-of-flight and
impact position on the trigger scintillators.

Particle identification was performed by comparing the response of the RT
scintillators to the energy loss, calculated using the formula of Bethe and
Bloch \cite{leo}. 
A particle identification parameter \pid\ was defined as
\begin{equation} \label{eq:pid}
        P_{id} = \frac{1}{N}\sum_{i=1}^{N} \frac{\limeas}{\licalc} 
\end{equation}
with $N$ the number of active RT scintillators and \limeas\ (\licalc)
the actual (calculated) response in the i{\em th} scintillator.
\pid\ will display a peak around 1 for deuterons, and a peak at smaller values
for protons and electrons.

For the kinematically overdetermined elastic scattering reaction, requiring
correlations between the scattering angles of the 
electron and the hadron reduces the number of protons even further. 

Figure \ref{fig:pid} shows the distribution of \pid\ and the coincidence
timing \taup between the scattered electron and the recoiling hadron. To
obtain this distribution $\pm$~2.5~$\sigma$ cuts were applied on their angular
correlations. A clear separation is observed between protons and deuterons. 
The proton contamination was estimated to be 4.6 \%. 
Analysis of a proton sample has shown that these 
have an analyzing power much lower than that of the deuterons. 
Therefore, the proton contamination is treated as an unpolarized background. 

In the event selection additional $\pm$~2.5~$\sigma$ cuts were applied on 
the coincidence time and on \pid. 
An asymmetry \atd was formed for events that fall within a \Q-bin,  
using the expression  
\begin{equation}\label{ttzucor}
\atd = \sqrt{2} \frac{N^+-N^-}{P_{zz}^+ N^- - P_{zz}^- N^+}
\end{equation}
with $N^+$ ($N^-$) the number of events in the \Q-bin considered when the 
target polarization was positive (negative). 
To correct for the fact that the
direction of the holding field -and thus the spin orientation- varies
over the length of the cell with respect to \qthree\, the uncorrected
tensor asymmetry \atd\ is weighted with \dtz\ from 
Eq. (\ref{eq:xsection}).  

Note from Eq. (\ref{eq:xsection}) that \atd\ contains small contributions
from \tto\ and \ttt.  Using the world data set for the unpolarized
structure functions \sfa\ and \sfb\ (see \cite{garcon} for an overview) it is
possible to correct for these contributions. For example,
\ttt\ can be expressed in terms of \sfa\ and \sfb\ directly as
$T_{22} = -\frac{\sqrt{3} \eta B}{8 (\eta + 1) ( A + B
\tan^2(\theta_e/2))}$ with $\eta = \frac{Q^2}{4M_{d}^2}$ and $M_d$ the
mass of the deuteron.
To investigate the sensitivity of the extraction procedure to the 
uncertainty in the input parameters (i.e. \Q, $\theta_e$, \dti,
\atd, \sfa\ and \sfb), these were varied independently within their error
and the extraction repeated. 
The total error was taken to be the quadratic sum of the separate errors.
Note that the main contribution to the systematic error in \atd\ comes from
the systematic uncertainty in the polarization. 

\begin{minipage}{234pt}
\begin{table}[tbp]
\centering
\caption{Result on \atd\ , \ttzst\ and  \fc\ with statistical and systematic
uncertainties, extracted from our \ttz\ measurements and the world
data on \sfa\ and \sfb. \label{tab:results}}
\medskip
\begin{tabular}{lccc}
\hline
Q [\fermi] & \atd & \ttzst(stat)(syst) & \fc(stat)(syst) \\
\hline
 2.03  & -0.683 & -0.713(0.082)(0.036) & 0.163(0.003)(0.014) \\   
 2.35  & -0.891 & -0.897(0.081)(0.045) & 0.100(0.003)(0.009) \\ 
 2.79  & -1.383 & -1.334(0.223)(0.066) & 0.035(0.015)(0.005) \\  
\hline
\end{tabular}
\end{table}
\end{minipage}

The observables \sfa, \sfb\ and \ttz\ provide three different
combinations of the form factors \fc, \fq\ and \fm, from which
these can be extracted. 
The result for \ttz\ was recalculated  
at $\theta_e = 70$\updeg, to allow a direct comparison
with the results of other experiments. 
The extracted values for \ttz\ and \fc\ are shown in Table
\ref{tab:results} and in Fig. \ref{fig:allttz}.  The new data on \ttz\
are each at least one $\sigma$ below the predictions of relativistic
models \cite{hummel,vanorden}. This confirms the findings of the
previous NIKHEF experiment \cite{eedmassi}.
 
To evaluate the model sensitivity of the \ttz\ data sets a \chis-analysis
was performed, for which the data measured
most recently at Bates \cite{garcon}, using a calibrated recoil 
polarimeter, and those from the NIKHEF experiments were selected. 
The data from BINP have poor accuracy at low \Q \cite{dmitriev} and poor
discriminating power in the \Q-range from 1 to 3 \fermi \cite{gilman}, since 
the \ttz\ values were extracted by normalizing one datum to a selected model 
prediction. The selected data sets are compared to the calculations of  
Wiringa \cite{wiringa}, Mosconi \cite{mosconi}, Hummel \cite{hummel}, 
Van~Orden \cite{vanorden} and Buchmann \cite{buchmann}.
The first two columns of Table \ref{tab:chittz} give the \chis-values
when only the \ttz\ data of either experiment are considered. In addition, 
both these experiments yielded data on other tensor analyzing powers: 
in the 95 data run of NIKHEF \cite{eedmassi} \ttt\ was also determined, 
and the Bates experiment
determined all tensor moments simultaneously. The last two columns of
the table give the results when all data are taken into account.

\begin{minipage}{234pt}
\begin{table}[htbp]
  \centering
    \caption{\chisred\ analysis for NIKHEF (N95+96: \protect\cite{eedmassi} 
    and present results) and Bates (B90: \protect\cite{garcon}) \tti\
	  data, against
    various model predictions. $N$ is the number of
    data points used in the analysis.\label{tab:chittz}}
  \begin{tabular}{lcccc}
    \hline
      &\multicolumn{1}{c}{\ttz(N95+96)} &\multicolumn{1}{c}{\ttz(B90)}
      &\multicolumn{1}{c}{\tti(N95+96)} &\multicolumn{1}{c}{\tti(B90)} \\
      &\multicolumn{1}{c}{N=4}          &\multicolumn{1}{c}{N=3}
      &\multicolumn{1}{c}{N=5}          &\multicolumn{1}{c}{N=9} \\
      \hline
      Wiringa        & 1.14 & 2.40 & 1.37  & 3.35 \\
      Mosconi           & 1.31 & 0.30 & 1.48  & 2.46 \\
      Hummel              & 2.80 & 0.89 & 2.68  & 2.54 \\
      Van~Orden         & 2.28 & 0.29 & 2.27  & 2.52 \\
      Buchmann          & 0.16 & 5.15 & 0.57  & 3.32 \\
          \hline
\end{tabular}
\end{table}
\end{minipage}

The two data sets lead to different conclusions about the
quality of the models. 
The NIKHEF set shows a preference for non-relativistic calculations with
realistic NN-potentials, when only the \ttz\ data are 
considered, and this conclusion remains unaltered when the datum on \ttt\ 
is included in the fit.    
The Bates data set, on the other hand, shows a preference for the
relativistic calculations, but loses most of its discriminating power
when all data on \tti\ are  taken into account, mainly due to an 
inconsistency in one value of \ttt.
The then available data on \ttz\ led Henning et al. \cite{henning} to
point out an inconsistency in the location of the minimum of the charge
form factor of two- and three-nucleon systems.

Stringent constraints are imposed on models by the extensive data set for
the unpolarized structure functions \sfa\ and \sfb, in addition to the 
polarized data.
In Table \ref{tab:alldat} the result of a \chis-analysis is given for \sfa\ 
and for \sfb, together with the overall \chis.
Data \cite{garcon} for \sfa\ and \sfb\ in the \Q-range of 
0.5 to 6.0  ~\fermi\ 
were taken into account. The normalization of each data set was varied 
within the quoted systematic uncertainty until a minimum value for the 
\chis\ was obtained.
The best description is given by the non-relativistic 
calculation of ref. \cite{wiringa} that includes the relevant corrections to
the impulse approximation, which conforms to the conclusions from the NIKHEF data.  
It should be noted that especially the inclusion of meson-exchange currents is
of great importance, both in describing the unpolarized and the polarized data. 

\begin{minipage}{234pt}
\begin{table}[htbp]
  \centering
        \caption{\chisred\ analysis of the \sfa\ and the \sfb\ world data set.
                  \label{tab:alldat}}\medskip
  \begin{tabular}{lccc}
    \hline
    Model       & 
      \multicolumn{1}{c}{\sfa}&
      \multicolumn{1}{c}{\sfb}&
      \multicolumn{1}{c}{\sfa+\sfb} \\
      &  
      \multicolumn{1}{c}{N=81}&
      \multicolumn{1}{c}{N=34}&
      \multicolumn{1}{c}{N=115}\\
    \hline
    Wiringa     &   5.6  & 5.9  &   5.7  \\
    Mosconi     &  11.0  & 1.8  &   8.3  \\
    Hummel      &  16.5  & 4.9  &  13.1  \\
    Van Orden   &  72.7  & 2.7  &  52.0  \\
    Buchmann    &  50.8  & 6.9  &  37.8  \\
    \hline
  \end{tabular} 
\end{table}
\end{minipage}

In conclusion, absolute measurements of the tensor analyzing power 
\ttz\ were performed in a \Q-range from 1.8 to 3.2 \fermi.  This new 
data set, together with that of a previous meassurement at NIKHEF, has
provided additional stringent constraints on the deuteron form factors.
Recently, an experiment \cite{kox} has been completed at Jefferson
Laboratory, which will provide accurate data on \ttz\ in a \Q-range from
4 to 6.5 \fermi, thus covering the expected position \cite{henning} of the
minimum in \fc.
Further measurements in a \Q-range of 2.1 to 4.4 \fermi\ will be performed
at NIKHEF, which will allow a direct comparison  with the MIT-Bates data.

This work was supported in part by the Stichting voor Fundamenteel 
Onderzoek der Materie (FOM), which is financially supported by the 
Nederlandse Organisatie voor Wetenschappelijk Onderzoek (NWO), the 
Swiss National Foundation, the National Science Foundation under 
Grants No. PHY-9316221 (Wisconsin), PHY-9200435 (Arizona State) and 
HRD-9154080 (Hampton), NWO Grant No. 713-119 and HCM Grants No. 
ERBCHBICT-930606 and ERB4001GT931472.

\newcommand{\PR}{{Phys. Rev. C}}
\newcommand{\PRL}{{Phys. Rev. Lett.}}
\newcommand{\PL}{{Phys. Lett.}}
\newcommand{\ZfP}{{Z. Phys.}}
\newcommand{\FBS}{{Few Body Syst.}}
\newcommand{\NIM}{{Nucl. Instrum. Methods}}
\newcommand{\NP}{{Nucl. Phys.}}
\newcommand{\AP}{{Ann. Phys.}}
\newcommand{\ARNPS}{{Ann. Rev. Nucl. Part.  Phys.}}
\newcommand{\CJP}{{Czech. J. Phys.}}
\newcommand{\MyJour}[5]{#1, #2\ \textbf{#3}\ (#4) #5}
\newcommand{\MyProc}[6]{#1,\ #2,\ #3,\ #4\ (#5), p. #6}
\newcommand{\MyBook}[4]{#1, {\em #2}, #3, (#4)}

\begin{figure}[p]
\caption{Particle identification parameter \pid\  (defined in the text)
versus coincidence time \taup. 
\label{fig:pid}
}
\end{figure}

\begin{figure}[p]
\caption{Extracted values (solid triangles) of \ttzst\ (top) and \fc\
(bottom) as a function of \Q\ compared to the world data and
selected calculations.  Data: solid triangles (present experiment), 
open squares \protect\cite{garcon}, solid square \protect\cite{eedmassi}, 
open diamond \protect\cite{dmitriev}, open triangles \protect\cite{gilman},
open circles \protect\cite{schulze} and open cross \protect\cite{popov}. 
Curves: short-dashed \protect\cite{wiringa}, dash-dotted \protect\cite{mosconi},
full \protect\cite{hummel}, long-dashed \protect\cite{vanorden}, 
dotted \protect\cite{buchmann}. The shaded 
area indicates the size of the systematic errors from the present experiment.
\label{fig:allttz}
}
\end{figure}

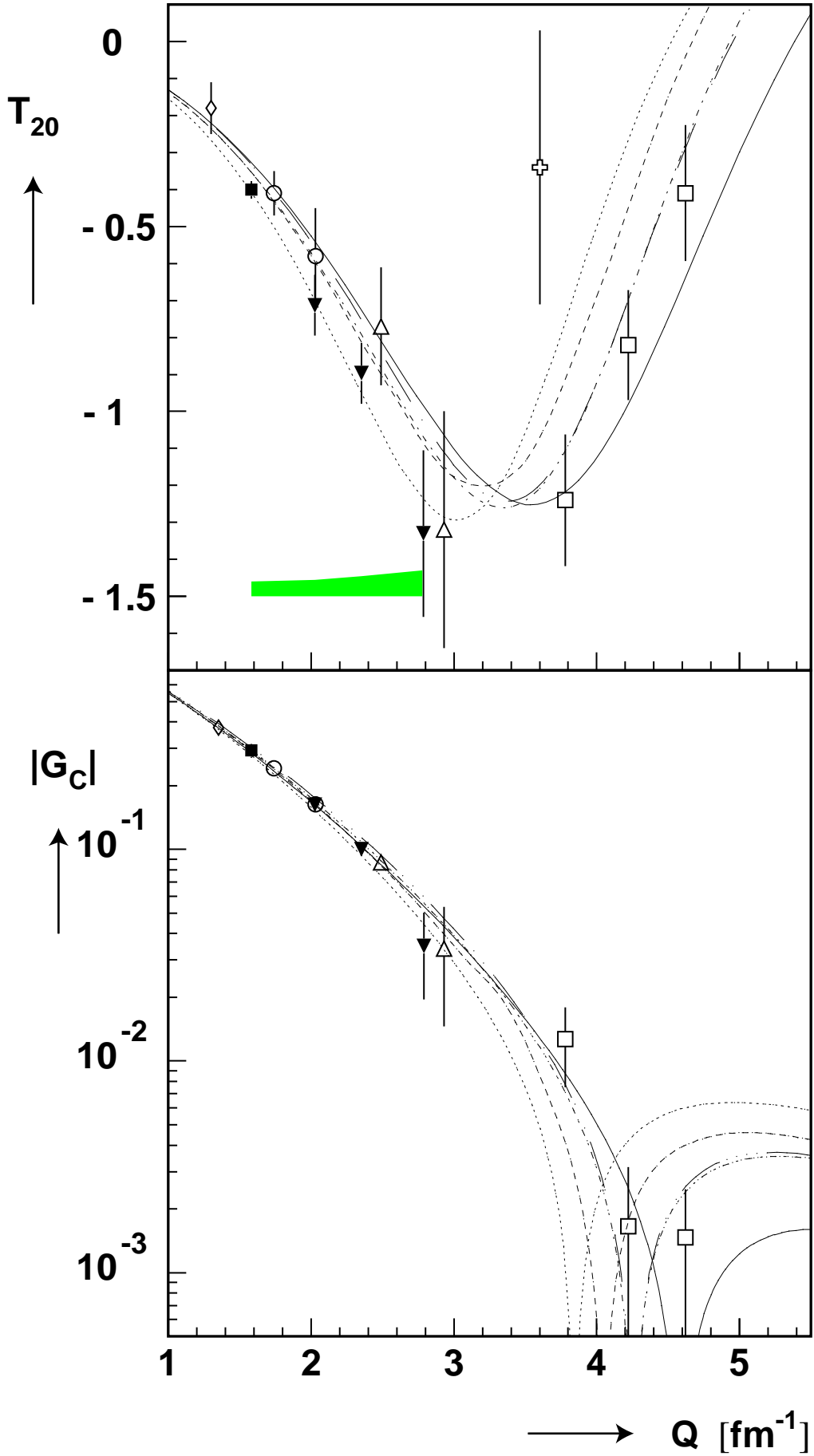
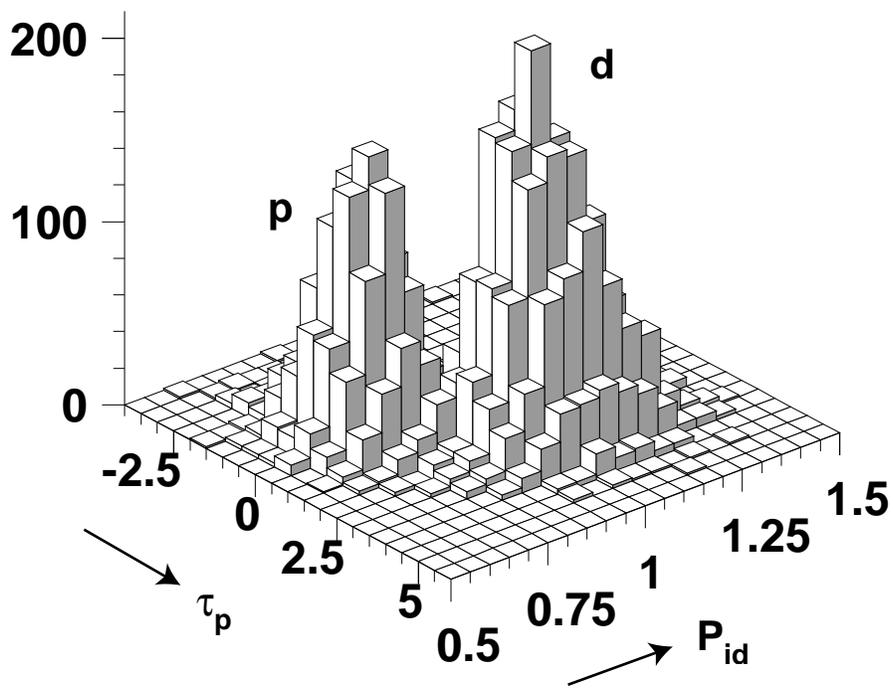
\end{document}